\documentclass[aps,pra,10pt,twocolumn,floatfix]{revtex4-1}
\usepackage{amsmath,amssymb,bm,graphicx}
\usepackage{times}

\usepackage{hyperref}
\hypersetup{
colorlinks=true,
linkcolor=blue,           
citecolor=blue,           
filecolor=magenta,     
urlcolor=cyan
}

\begin{document} 

\title{Emergence of continuous rotational symmetries in ultracold atoms coupled to optical cavities}
\author{E.~I.~Rodr\'iguez Chiacchio and A.~Nunnenkamp}
\affiliation{Cavendish Laboratory, University of Cambridge, Cambridge CB3 0HE, United Kingdom}

\date{\today}

\begin{abstract}
We investigate the physics of a gas of ultracold atoms coupled to three single-mode optical cavities and transversely pumped with a laser. Recent work has demonstrated that, for two cavities, the $\mathbb{Z}_{2}$ symmetries of each cavity can be combined into a global $U(1)$ symmetry \cite{Esslinger3}. Here, we show that when adding an extra cavity mode, the low-energy description of this system can additionally exhibit an $SO(3)$ rotational symmetry which can be spontaneously broken. This leads to a superradiant phase transition in all the cavities simultaneously, and the appearance of Goldstone and amplitude modes in the excitation spectrum. We determine the phase diagram of the system, which shows the emergence and breaking of the continuous symmetries and displays first- and second-order phase transitions. We also obtain the excitation spectrum for each phase and discuss the atomic self-organized structures that emerge in the different superradiant phases. We argue that coupling the atoms equally to $n$ different modes will in general generate a global $SO(n)$ symmetry if the mode frequencies can be tuned to the same value.
\end{abstract}

\maketitle

\section{Introduction}

Ultracold atomic gases constitute one of the most versatile platforms for the quantum simulation of many-body physics \cite{RevBloch,RevBloch2012}. In recent years, the scope of these systems has been greatly enhanced by considering interactions between atoms and dynamical light fields, generated by optical cavities \cite{Review}.

A paradigmatic example of this is the experimental realization of the well-known Dicke model \cite{Dicke} with a gas of ultracold atoms loaded inside a single-mode optical cavity \cite{Esslinger,HemmerichPRL2014}. In this setup, the system undergoes a superradiant phase transition, associated with the breaking of a $\mathbb{Z}_{2}$ symmetry \cite{Emary}, where the cavity field becomes macroscopically occupied and the atoms self-organize in a checkerboard pattern \cite{DomokosPRL2002,DomokosPRL2010,Bhaseen1,Bhaseen2}. The high degree of control characterizing these experimental systems allowed for the study of the symmetry breaking process \cite{EsslingerDickeSB}, measurement of the excitation spectrum \cite{EsslingerDickeRoton}, and real-time observation of the fluctuations, using photon loss processes to perform in-situ montoring of the system \cite{EsslingerDickeFluct,HemmerichPNAS2014}.

Recent work brought these concepts to a higher level of complexity by considering the effects of coupling a second cavity to the atomic cloud \cite{Esslinger3}. The combined system inherits a $\mathbb{Z}_{2} \times \mathbb{Z}_{2}$ symmetry, which can be broken independently, yielding a superradiant state in one of the cavities. More importantly, it was found that when the cavities are coupled symmetrically to the atoms, the system exhibits an overall continuous $U(1)$ symmetry, which upon breaking, leads to the presence of superradiant emission in both cavities simultaneously. These results were corroborated by the observation of the associated Goldstone and Higgs modes \cite{Esslinger4}. Further studies have also considered the robustness of this symmetry \cite{PiazzaZwergerNJP} and the effects of inter-cavity photon scattering processes on the ground state phase diagram \cite{EsslingerOrdPar,GopalakrishnanPRA2017}.

Symmetry enhancement of this type was previously discussed in the context of circuit QED \cite{BaksicCiutiPRL} and for atoms coupled to two-mode cavities \cite{FanPRA2014}, which has recently been analyzed in detail for generic atom-light couplings and including the effects of photon loss \cite{KeelingPRA2018}. These systems exhibit complex ground and steady state phase diagrams, including multicritical points, and qualitatively different phases, resulting from  the different underlying symmetries and their spontaneous breaking. With the emergence of these rich phenomena, it is intriguing to ask what other symmetries can arise for atom-cavity systems when further increasing their complexity, given their strong potential as quantum simulators.
\begin{figure}[t]
\centering
\includegraphics[width=0.7\columnwidth]{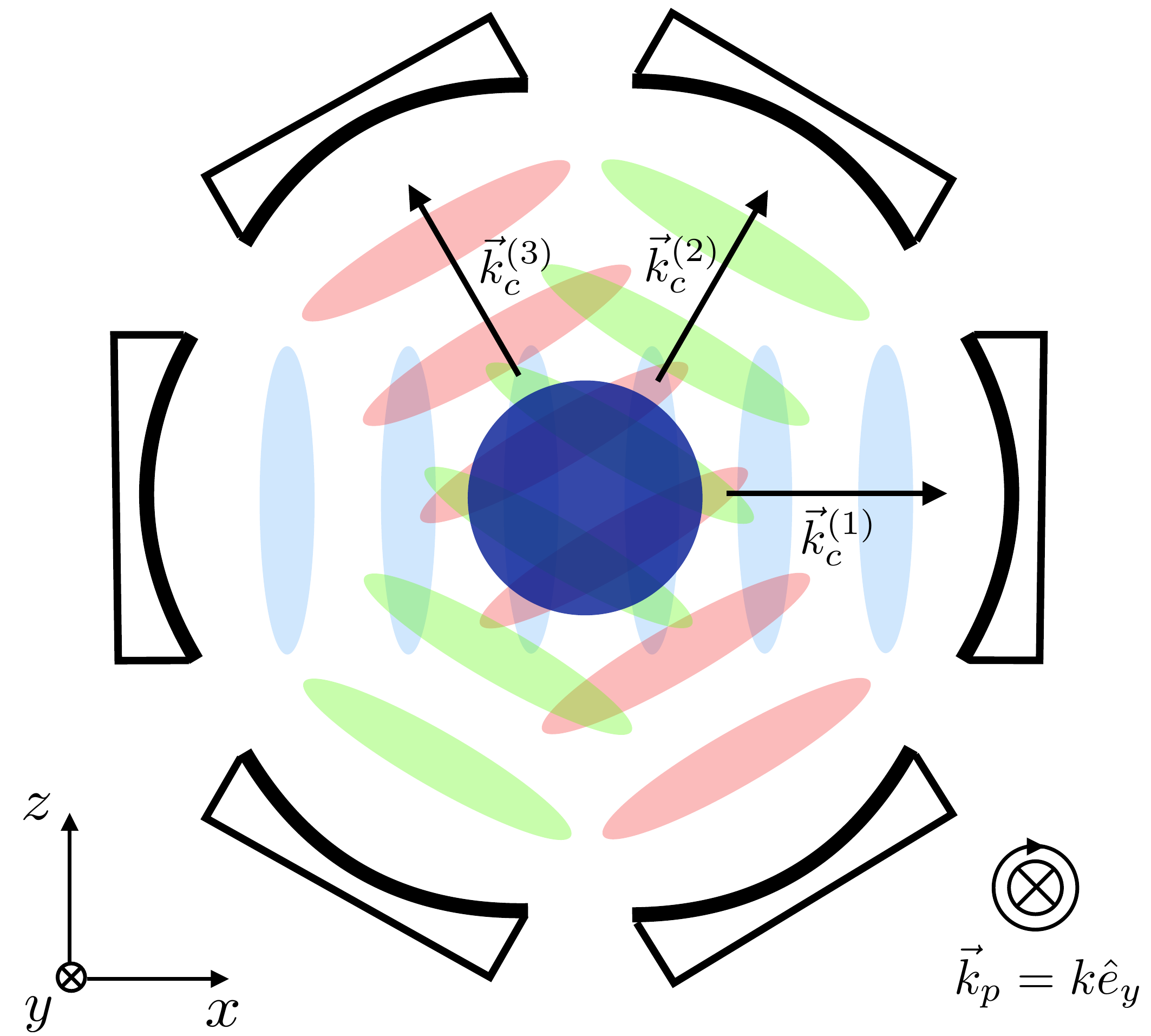}
\caption{\label{FigModel} A gas of ultracold atoms (blue circle) is placed inside three high-finesse cavities all in the $x$-$z$ plane and aligned at the same angle from each other. The system is transversely pumped by a circularly polarized laser in the $y$ direction.}
\end{figure}

In this paper, we explore the consequences of adding a third cavity mode into a setup similar to \cite{Esslinger3}. The effective model is presented in Sec.~\ref{Model}. We show that the system possesses a $\mathbb{Z}_{2} \times  \mathbb{Z}_{2} \times \mathbb{Z}_{2}$ symmetry, which cannot only be combined to form the previously observed $U(1)$ symmetry, but also yields a global $SO(3)$ rotational symmetry when all cavities are coupled symmetrically to the atomic cloud (Sec.~\ref{Symmetries}). We additionally find that this generalizes to an $SO(n)$ symmetry when the atoms are symmetrically coupled to $n$ cavities. We obtain the ground state phase diagram of the system using a mean-field approximation and characterize the emergent phases resulting from the spontaneous breaking of the different symmetries (Sec.~\ref{MFPD}). We complement this analysis by calculating the excitation spectrum and studying its behavior when crossing the different critical points present in the system (Sec.~\ref{ExcSpec}). In Sec.~\ref{SO}, we discuss the possible self-organized structures of the atoms in the different phases. We conclude in Sec.~\ref{Conc}.

\section{Model}
\label{Model}
We consider a system of three intersecting single-mode optical cavities, symmetrically aligned in the $x$-$z$ plane, with a gas of ultracold atoms forming a Bose-Einstein condensate (BEC) placed in the center, see Fig.~\ref{FigModel}. The system is additionally pumped by a circularly polarized laser in the $y$ direction, which is reflected off a mirror (not shown in Fig.~\ref{FigModel}), generating a standing wave potential for the atoms. Two-photon scattering processes between the pump and the cavities mediate momentum transitions for the atoms from the $|\vec{k}_{0}\rangle$ BEC state into a set of twelve excited states $|\vec{k}^{(i)}=\pm(\vec{k}_{p} \pm \vec{k}^{(i)}_{c}) \rangle$, with $\vec{k}_{p}$ and $\vec{k}_{c}^{(i)}$ the wave vectors of the pump and cavity $i$, respectively (see Fig.~\ref{FigModel}). For the case $|\vec{k}_{c}^{(i)}|=|\vec{k}_{p}|=k$, these excited states become degenerate, yielding a low-energy description which in the rotating frame of the pump reads [see Appendix \ref{AppA} for details] ($\hbar=1$)
\begin{equation}
\label{eq:er0002}
 \hat{H}= \sum_{i=1}^{3}(-\Delta_{i}) \hat{a}_{i}^{\dagger} \hat{a}_{i} + \omega \hat{b}^{\dagger}_{i} \hat{b}_{i} + \frac{\lambda_{i}}{\sqrt{N}} (\hat{a}_{i}^{\dagger}+ \hat{a}_{i}) \left( \hat{b}^{\dagger}_{i} \hat{b}_{0} + \hat{b}_{0}^{\dagger} \hat{b}_{i} \right)  ,
\end{equation}
where $\hat{a}_{i}$ is the annihilation operator for photons in cavity $i$, with $\Delta_{i}=\omega_{p}-\omega_{c}$ the cavity-pump detuning, $\hat{b}_{0}$ and $\hat{b}_{i}$ are bosonic annihilation operators for atoms in the $|\vec{k}_{0}\rangle$ and the $|\vec{k}^{(i)}\rangle$ states, respectively, and $\omega$ is the energy difference between $|\vec{k}_{0}\rangle$ and $|\vec{k}^{(i)}\rangle$. The interaction term corresponds to transitions between $|\vec{k}_{0}\rangle$ and $|\vec{k}^{(i)}\rangle$ mediated by the emission or absorption of a photon in cavity $i$, with strength $\lambda_{i}/\sqrt{N}$, where $N$ is the total number of atoms in the system. We focus on the case where $\Delta_{i}=\Delta<0$ for all $i$, and use the coupling strengths $\lambda_{i}$ as control parameters.

\section{Symmetries}
\label{Symmetries}
For general $\lambda_{i}$, the Hamiltonian \eqref{eq:er0002} possesses a $\mathbb{Z}_2 \times\mathbb{Z}_2 \times\mathbb{Z}_2 $ symmetry, associated with parity transformations of the form
\begin{equation}
\label{eq:er0003}
\left(\hat{a}_{i},\hat{a}_{i}^{\dagger},\hat{b}_{i},\hat{b}_{i}^{\dagger} \right) \longrightarrow -\left(\hat{a}_{i},\hat{a}_{i}^{\dagger},\hat{b}_{i},\hat{b}_{i}^{\dagger} \right) ,
\end{equation}
for $i=1,2,3$. If two of the cavities have the same coupling strength $\lambda_{1}=\lambda_{2}=\lambda \neq \lambda_{3}$, their corresponding $\mathbb{Z}_{2} \times \mathbb{Z}_{2}$ is combined into an $U(1)$ symmetry associated with rotations between the degrees of freedom cavities 1 and 2
\begin{equation}
\label{eq:er0004}
\begin{pmatrix} \hat{a}_{1} \\ \hat{a}_{2} \end{pmatrix} \rightarrow \hat{\mathcal{R}}_{\theta} \begin{pmatrix} \hat{a}_{1} \\ \hat{a}_{2} \end{pmatrix}, \qquad \begin{pmatrix} \hat{b}_{1} \\ \hat{b}_{2} \end{pmatrix} \rightarrow \hat{\mathcal{R}}_{\theta} \begin{pmatrix} \hat{b}_{1} \\ \hat{b}_{2} \end{pmatrix},
\end{equation}
with
\begin{equation}
\label{eq:er0005}
\hat{\mathcal{R}}_{\theta} = \begin{pmatrix} \cos{\theta} & -\sin{\theta} \\ \sin{\theta} & \cos{\theta} \end{pmatrix} .
\end{equation}
The overall symmetry of the system then becomes $U(1) \times \mathbb{Z}_2$. Lastly, for the case $\lambda_{i}=\lambda$ for all $i$, the Hamiltonian becomes invariant under the transformation $\hat{\mathcal{R}}_{\theta}$ acting on any pair of cavities. For three cavities, we can associate this invariance a global $SO(3)$ symmetry, corresponding to the three possible rotations between cavities.

By tuning the coupling strengths, we can then interpolate between the regimes where the system acquires different symmetries, which can be spontaneously broken separately. Breaking of a discrete $\mathbb{Z}_{2}$ symmetry is associated with the system undergoing a superradiant phase transition, where one of the cavities acquires a macroscopic occupation number, accompanied by a self-organization of the atomic cloud in a checkerboard pattern, resulting from the interference between the pump and the macroscopic cavity field \cite{Esslinger,DomokosPRL2002,DomokosEPJ2008}. In the case of a $U(1)$ symmetry breaking, the macroscopic light field is arbitrarily spread between the two symmetrical cavities \cite{Esslinger3}, consequence of the ground state degeneracy, and the self-organization pattern is then given by the interference between the pump and the two cavity fields. As discussed in the next section, the same occurs when the emergent $SO(3)$ symmetry is broken, but with the light field spread among the three cavities instead.

In the following, we make these notions precise by studying the phase diagram using a mean-field approach, obtaining the excitation spectrum, and analyzing the self-organization of the atoms due to the light-field interference.

\section{Mean-field phase diagram}
\label{MFPD}
We obtain the ground state phase diagram making use of the mean-field approximation which is valid in the thermodynamic limit $N \rightarrow \infty$. We start by introducing the order parameters $\alpha_{i} = \langle \hat{a}_{i} \rangle$ and $\beta_{i} = \langle \hat{b}_{i} \rangle$ into Eq.~\eqref{eq:er0002}
\begin{align}
\label{eq:er0008}
E_{\textrm{MF}} = \sum_{i=1}^3 & (-\Delta) |\alpha_{i}|^2 + \omega |\beta_{i}|^2 + \frac{\lambda_{i}}{\sqrt{N}} \left( \alpha_{i} + \alpha_{i}^{*} \right) \times \nonumber \\
&\left( \beta_{i} + \beta_{i}^{*} \right) \sqrt{N-\sum_{j} |\beta_{j}|^2} ,
\end{align}
where we made use of particle number conservation $|\langle \hat{b}_{0} \rangle |^2 = N - \sum_{j=1}^3 |\beta_{j}|^2$. We reduce the degrees of freedom by minimizing the energy with respect to the cavity field, leading to $\alpha_{i} = -\frac{\lambda_{i}}{(-\Delta)\sqrt{N}}\left( \beta_{i} + \beta_{i}^{*} \right) \sqrt{N-\sum_{j} |\beta_{j}|^2}$ and 
\begin{equation}
\label{eq:er0009}
E_{\textrm{MF}} = \sum_{i=1}^3 \omega |\beta_{i}|^2 - \lambda_{i}^2\frac{\left( \beta_{i} + \beta_{i}^{*} \right)^2}{(-\Delta)N} \left( N-\sum_{j} |\beta_{j}|^2 \right).
\end{equation}
From $\beta_{i}= \textrm{Re}[\beta_{i}] + i\textrm{Im}[\beta_{i}]$ and minimizing $E_{\textrm{MF}}$ with respect to $\textrm{Im}[\beta_{i}]$, we obtain $\textrm{Im}[\beta_{i}]=0$, resulting in
\begin{equation}
\label{eq:er0010}
E_{\textrm{MF}} =  \omega \left( \sum_{i=1}^3 \frac{\mu_{i}-1}{\mu_{i}} \beta_{i}^2 - \frac{1}{N}\sum_{i,j=1}^3 \frac{\beta_{i}^2 \beta_{j}^2}{ \mu_{i}}  \right),
\end{equation}
with $\mu_{i}=\lambda_{\textrm{cr}}^2/\lambda^2_{i}$, where $\lambda_{\textrm{cr}}=\frac{\sqrt{(-\Delta)\omega}}{2}$ is the critical coupling strength. The ground state phase diagram of the system then follows from the global minima of $E_{\textrm{MF}}$ as a function of the parameters $\lambda_{i}$. This yields four different phases, which we denote as normal (NP), single cavity superradiant (S1), double cavity superradiant (S2) and triple cavity superradiant (S3) [see Fig.~\ref{gsmfpd}]:
\begin{figure}[t]
\centering
\includegraphics[width=0.8\columnwidth]{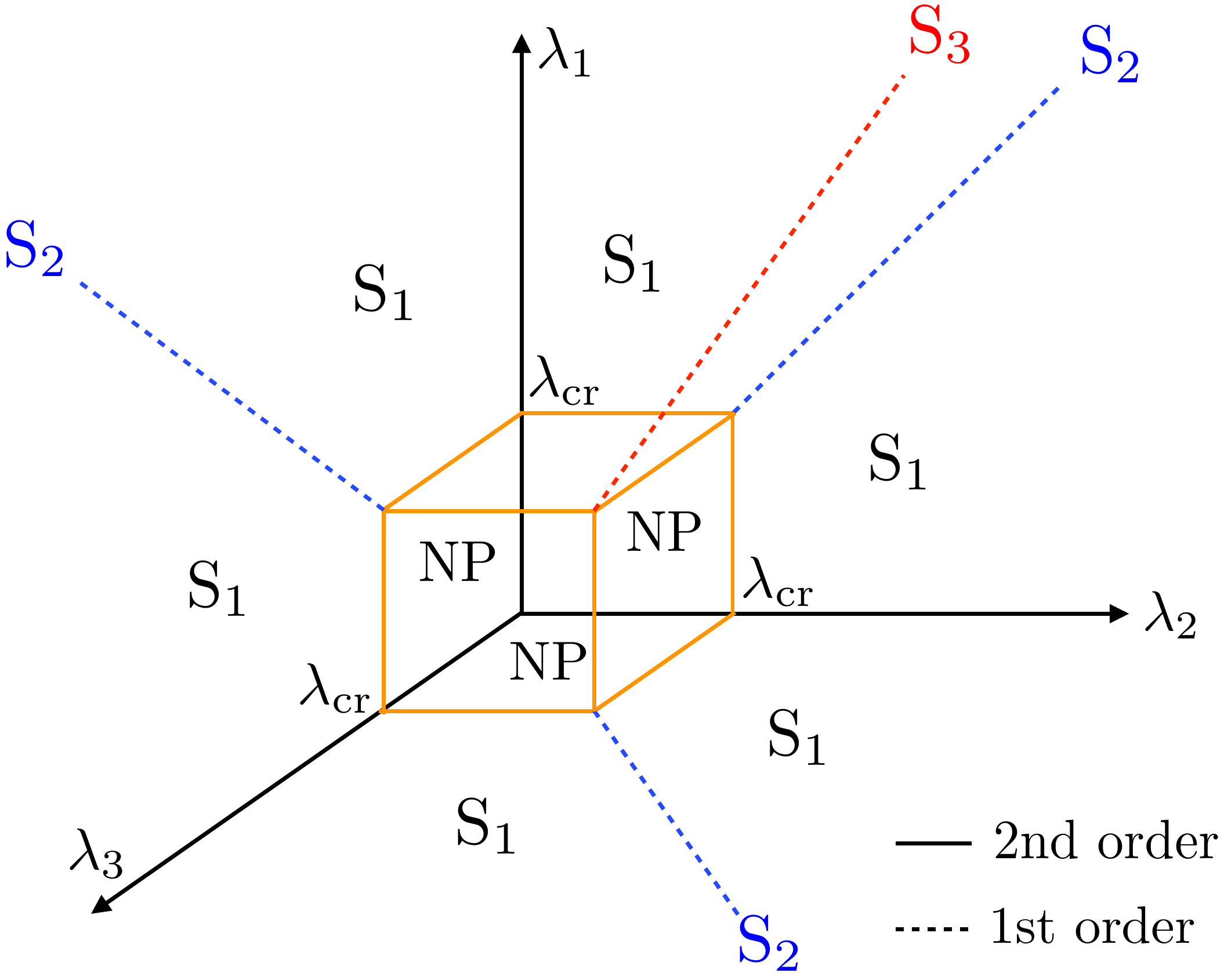}
\caption{\label{gsmfpd} Mean-field ground state phase diagram of the system as a function of the atom-light couplings $\lambda_{i}$, resulting from the minima of $E_{\textrm{MF}}$ \eqref{eq:er0010}. The orange cubic region where all $\lambda_{i}<\lambda_{\textrm{c}}$ corresponds to the normal phase (NP). The regions with $\lambda_{i}>\lambda_{\textrm{cr}}$ and $\lambda_{i}>(\lambda_{j},\lambda_{k})$ are associated with single cavity superradiance (S1). Blue lines denote the edge of the planes with $\lambda_{j}=\lambda_{k}=\lambda>\lambda_{\textrm{cr}}$ and $\lambda>\lambda_{i}$, where the broken $U(1)$ symmetry yields superradiance in two cavities (S2). The diagonal red line denotes the region $\lambda_{i}=\lambda>\lambda_{\textrm{cr}}$, for all $i$, where the spontaneous SO(3) symmetry breaking leads to superradiance in all cavities (S3). Solid lines correspond to second-order phase transitions and dashed ones to first-order.}
\end{figure}
\begin{itemize}
\item (NP) - For all $\lambda_{i}<\lambda_{\textrm{cr}}$, the only minimum of $E_{\textrm{MF}}$ is the trivial solution, $\beta_{i}=0$ for all $i$, where there is no macroscopic occupation in any cavity and the atoms remain in the BEC state.
\item (S1) - For $\lambda_{i}>\lambda_{\textrm{cr}}$  and $\lambda_{i} > (\lambda_{j},\lambda_{k})$, one of the $\mathbb{Z}_{2}$ symmetries is spontaneously broken and the energy develops two minima at $\beta_{i}=\pm \sqrt{\frac{N}{2}(1-\mu_{i})}$, $\beta_{j}=\beta_{k}=0$, corresponding to the two possible self-organized patterns and the presence of a macroscopic light field in cavity $i$.
\item (S2) - For $\lambda_{i}=\lambda_{j}=\lambda>\lambda_{\textrm{cr}}$ and $\lambda>\lambda_{k}$, the minima of $E_{\textrm{MF}}$ correspond to a circumference in the $\beta_{i}$-$\beta_{j}$ plane, parametrized by 
\begin{equation}
\label{eq:er0011a}
\begin{pmatrix} \beta_{i} \\ \beta_{j} \end{pmatrix} = \sqrt{\frac{N}{2}(1-\mu)} \begin{pmatrix} \cos{\theta} \\ \sin{\theta} \end{pmatrix} ,
\end{equation}
with $\beta_{k}=0$. This corresponds to the spontaneous breaking of the continuous $U(1)$ symmetry, where two cavities become superradiant and the relative distribution of light intensity is given by the angle $\theta$. 
\item (S3) - For all $\lambda_{i}=\lambda>\lambda_{\textrm{cr}}$, the energy minima span a spherical surface in order parameter space, meaning that all modes become macroscopically occupied
\begin{equation}
\label{eq:er0011b}
\begin{pmatrix} \beta_{1} \\ \beta_{2} \\ \beta_{3} \end{pmatrix} = \sqrt{\frac{N}{2}(1-\mu)} \begin{pmatrix} \sin{\phi}\cos{\theta} \\ \sin{\phi}\sin{\theta} \\ \cos{\phi}  \end{pmatrix}.
\end{equation}
This triple superradiant phase emerges from the spontaneous breaking of the $SO(3)$ symmetry, where the two angles $(\phi,\theta)$ parametrize the distribution of light intensity.
\end{itemize}

It is important to note that for all superradiant phases the field intensity $\beta^2=\sum_{i} \beta_{i}^2$ is the same. Physically, this results from it being fixed only by the driving strength, encoded in $\lambda_{i}$, and the total number of atoms in the system. Therefore, the ground states of a specific superradiant phase are connected by transformations that preserve the value of $\beta^{2}$. Geometrically, in $n>1$ dimensions, such transformations correspond to proper rotations (and parity transformations in the single cavity case $n=1$),  which are nothing but the generators of the $SO(n)$ group. Thus, for the generic case of symmetrical coupling of the atoms to $n$ different cavities, this results in an overall $SO(n)$ symmetry. We can then understand the emergence of rotational symmetries in the system as the ground states of a superradiant phase conserving the total field intensity, and being related to each other only by a redistribution of the intensity among the different cavity modes. 
\begin{figure}[t]
\centering
\includegraphics[width=\columnwidth]{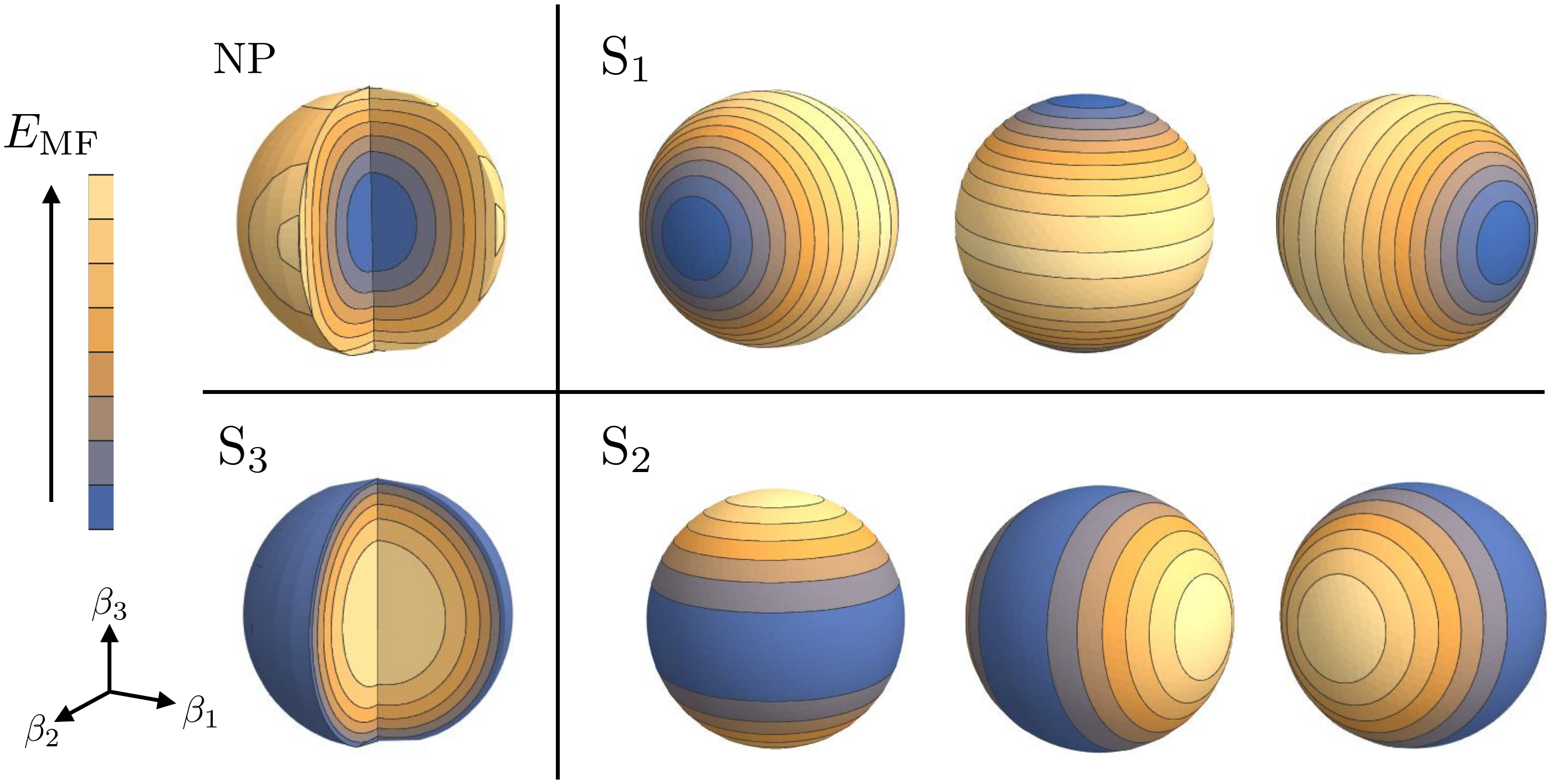}
\caption{\label{GSsol} The mean-field energy \eqref{eq:er0010} as a function of the order parameters for the four different phases. The energy minimum (blue region) sits at the origin in the NP, and its position  becomes non-vanishing in the ordered phases. In the S1 phase, the three solutions correspond to one of the cavities becoming superradiant, and for the S2 phase to one pair of cavities becoming superradiant simultaneously, where different points in the ground state manifold are associated with different intensity distributions between the cavities. In the S3 phase, all cavities become superradiant and the ground state manifold corresponds to a spherical surface where the different ground states can also be associated with different intensity distributions.}
\end{figure}

In Fig.~\ref{GSsol}, we show the ground state manifold as a function of the order parameters for the four different phases. From the form of $E_{\textrm{MF}}$, we find that all transitions from the normal phase into any of the superradiant ones are of second order, as the minimum smoothly shifts away from the trivial solution when the system crosses the phase boundary. In contrast, transitions among superradiant phases are of first order since the order parameters change values discontinuously at the critical points.

\section{Excitation spectrum}
\label{ExcSpec}

\begin{figure}[t]
\centering
\includegraphics[width=\columnwidth]{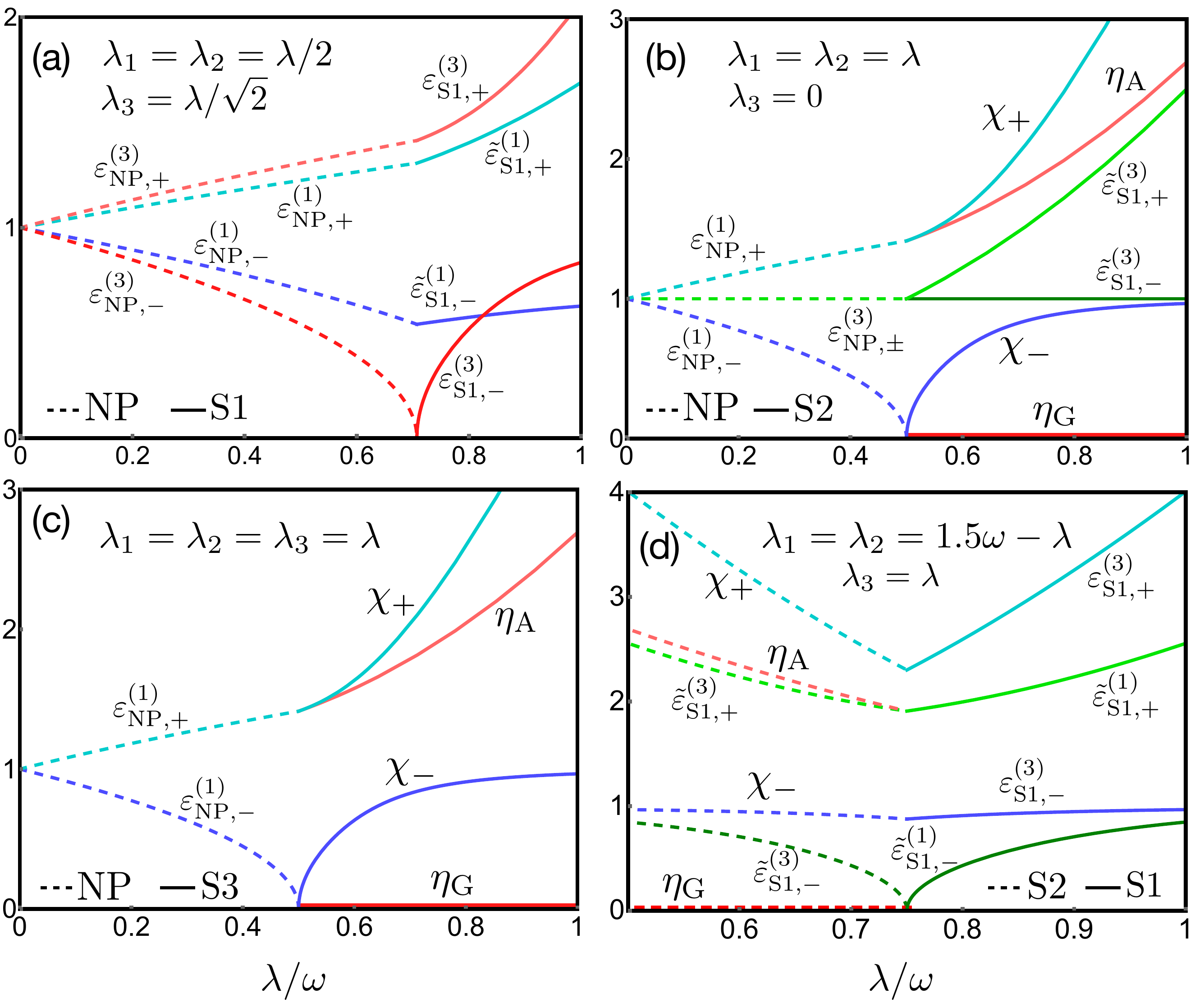}
\caption{\label{PlotExc} Excitation spectrum (in units of $\omega$) when crossing the transitions from: (a) NP to S1, (b) NP to S2, (c) NP to S3 and (d) S2 to S1 (first order). In (b) and (c) we observe the emergence of Goldstone modes, resulting from the spontaneous breaking of the continuous symmetries $U(1)$ and $SO(3)$, respectively. In all cases $\omega=-\Delta$.}
\end{figure}

Following the methods in Refs.~\cite{Emary,BaksicCiutiPRL}, the spectrum of fluctuations can be obtained by displacing the operators in Eq.~\eqref{eq:er0002} by their mean value, i.e.~$\hat{a}_{i} \rightarrow \hat{c}_{i} + \alpha_{i}$, $\hat{b}_{i} \rightarrow \hat{d}_{i} - \beta_{i}$, and neglecting terms of order $\mathcal{O}(\frac{1}{N})$. This yields a bilinear Hamiltonian in the fluctuations ($\hat{c}_{i}$,$\hat{d}_{i}$) from which the spectrum can be extracted using a Bogoliubov transformation, see Appendix \ref{AppB} for details. In the normal phase and the S1 phase, the spectrum reads
\begin{align}
\label{eq:er0012a}
\varepsilon_{\textrm{NP},\pm}^{(i)2} &= \frac{1}{2} \left[ \Delta^2 + \omega^2 \pm \sqrt{(\Delta^2-\omega^2)^2 - 16\lambda_{i}^2\Delta\omega } \right] \\
\label{eq:er0012b}\varepsilon_{\textrm{S1},\pm}^{(i)2} &= \frac{1}{2} \left[ \frac{\omega^2}{\mu_{i}^{2}} + \Delta^{2} \pm \sqrt{\left( \frac{\omega^2}{\mu_{i}^2} - \Delta^2 \right)^2 + 4\omega^2 \Delta^2} \right] \\
\label{eq:er0012c}\tilde{\varepsilon}_{\textrm{S1},\pm}^{(j \neq i),2} &= \frac{1}{2} \Bigg[ \Delta^2 + \frac{\omega^2}{4\mu_{i}^2}(1+\mu_{i})^2 \pm \nonumber \\
 & \sqrt{\left(\Delta^2-\frac{\omega^2}{4\mu_{i}^2}(1+\mu_{i})^2\right)^2 + 4\lambda_{j}^2\Delta \frac{\omega}{\mu_{i}}(1+\mu_{i})^2 } \Bigg], 
\end{align}
where we considered cavity $i$ to be in the superradiant state. These excitations are shown in Fig.~\ref{PlotExc}(a), where we observe how the lowest energy branch vanishes at the critical point between the NP and the S1 phases, to increase again in the S1 phase, as expected from the spontaneous breaking of a discrete symmetry. The dependence of \eqref{eq:er0012c} in $\lambda_{i}$, through $\mu_{i}=\lambda^{2}_{\textrm{cr}}/\lambda^{2}_{i}$, stems from the transition boundaries to other superradiant phases also being dependent on $\lambda_{i}$ (blue lines in Fig.~\ref{gsmfpd}). In the S2 phase, the excitation branches of the cavity that remains in the normal phase correspond to $\tilde{\varepsilon}_{\textrm{S1},\pm}^{(j)}$, whereas the superradiant branches mix, yielding
\begin{align}
\label{eq:er0012d}\eta_{\textrm{G}}^2&=0 \\
\label{eq:er0012e} \eta^2_{\textrm{A}}&=\frac{1}{4\Delta^2}\left( 4\Delta^4+16\lambda^4-8\lambda^2\Delta\omega+\Delta^2\omega^2 \right) \\
\label{eq:er0012f}\chi_{\pm}^2&=\frac{1}{2\Delta^2}\left( \Delta^{4}+16\lambda^{4} \pm \sqrt{(\Delta^4-16\lambda^4)^2 + 4\Delta^{6}\omega^2} \right) ,
\end{align}
where $\eta_{\textrm{G},\textrm{A}}$ correspond to the Goldstone and amplitude modes, respectively, associated with the breaking of the continuous $U(1)$ symmetry. This can be observed in Fig.~\ref{PlotExc}(b), with the appearance of a vanishing mode ($\eta_{G}$) as the gap closes at the critical point. One can also see how the excitations in \eqref{eq:er0012d}-\eqref{eq:er0012f}, result from the mixing of the superradiant cavity modes, as the excitations for the non-superradiant cavity (solid green lines) emerge from the branches of the same cavity in the normal phase (dashed green lines). For the S3 phase, the mode mixing leads to the same spectrum of the S2 phase, where now the $\eta_{\textrm{G},\textrm{A}}$ modes become double degenerate instead. This is shown in Fig.~\ref{PlotExc}(c), where the excitations associated with a non-superradiant cavity in (b) are not present any more. The increase in the number of Goldstone modes follows from the spontaneous breaking of a higher dimensional symmetry, namely $SO(3)$, and can be understood as excitations along the angular directions of the ground state manifold. In Fig.~\ref{PlotExc}(d) we also include the excitations across a transverse cut in the phase diagram, where a first-order phase transition occurs when going from the S2 into the S1 phase.

\section{Atomic self-organization}
\label{SO}
The presence of superradiance in the system is accompanied by atomic self-organization, where the atoms sit at the minima of the effective potential generated by the interference between the pump and the cavity light fields. In this section, we present the self-organized patterns that arise for the different superradiant phases. The total effective potential reads
\begin{figure}[t]
\centering
\includegraphics[width=\columnwidth]{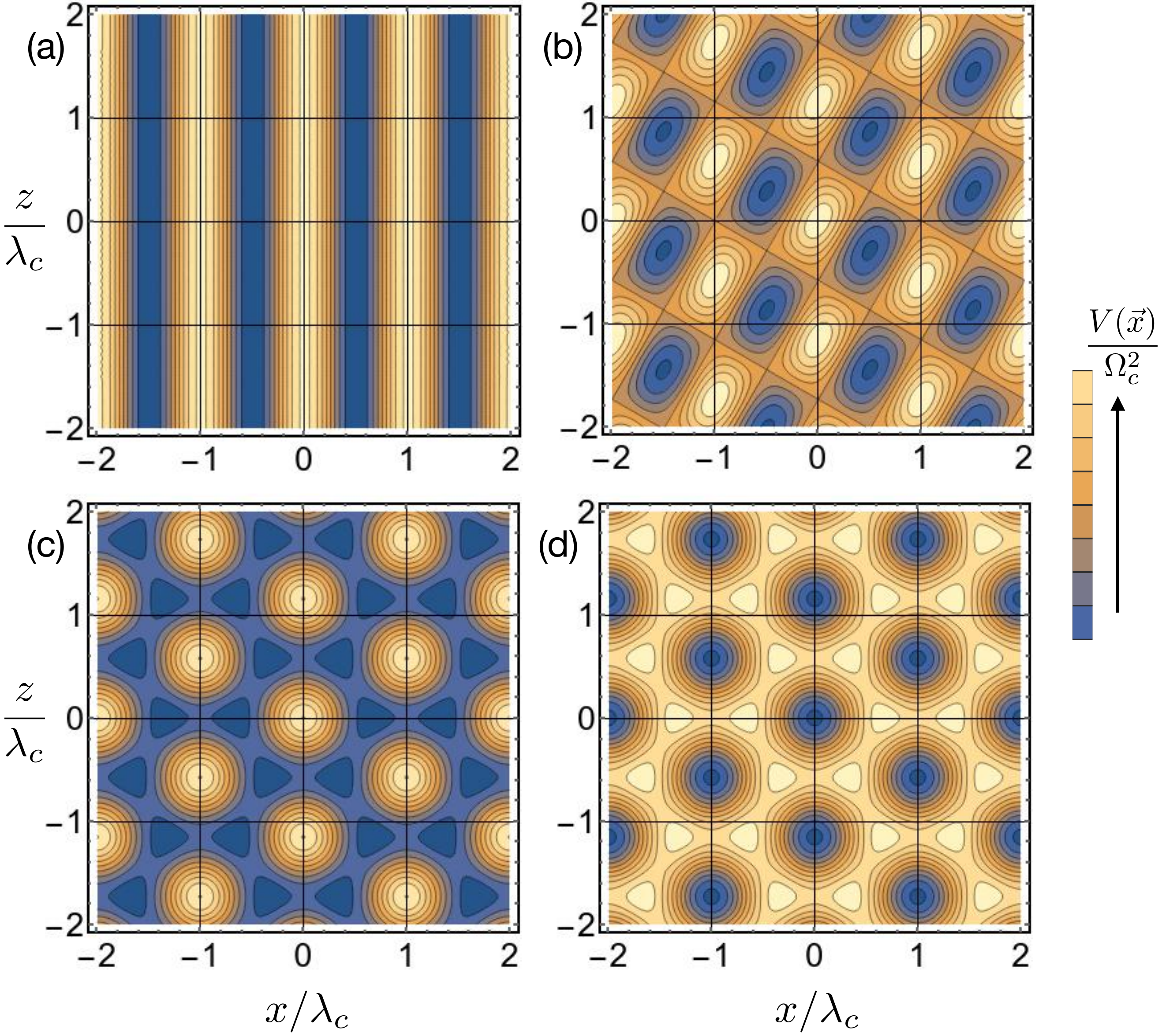}
\caption{\label{plotSO}Two-dimensional cuts ($y=0$) of the effective potential $V(\vec{x})$ generated by the light fields for $\Omega_{p}=10\Omega_{c}$, in the S3 phase corresponding to: (a) $\phi=\pi/2$, $\theta=0$ (equivalent to the S1 phase); (b) $\phi=\pi/4$, $\theta=0$ (equivalent to the S2 phase); (c) $\phi=\arctan{(\sqrt{2})}$, $\theta=\pi/4$; and (d) $\phi=\pi - \arctan{(\sqrt{2})}$, $\theta=3\pi/4$.}
\end{figure}
\begin{equation}
\label{eq:er0013}
V(\vec{x})=\left[ \Omega_{p} \cos{(ky)} + \sum_{i=1}^{3} \Omega_{i}\cos{(\vec{k}_{c}^{(i)}\vec{x})} \right]^{2},
\end{equation}
where $\Omega_{(p,i)}=\sqrt{U_{(p,i)}}$ are the pump and cavity amplitudes, respectively, with $U_{(p,i)}$ the potential depths (see Appendix \ref{AppA}), and the cavity wave vectors are defined as $\vec{k}_{c}^{(1)}=k \hat{e}_{x}$, $\vec{k}_{c}^{(2)}=\frac{k}{2}(\hat{e}_{x}+ \sqrt{3} \hat{e}_{z})$ and $\vec{k}_{c}^{(3)}=\frac{k}{2}(\hat{e}_{x}- \sqrt{3} \hat{e}_{z})$, as shown in Fig.~\ref{FigModel}. In Fig.~\ref{plotSO}, we present four two-dimensional cuts of $V(\vec{x})$, for the self-organized atoms in the S3 phase, where the amplitudes are distributed according to $\Omega_{1}=\Omega_{c}\sin{\phi}\cos{\theta}$, $\Omega_{2}=\Omega_{c}\sin{\phi}\sin{\theta}$ and $\Omega_{3}=\Omega_{c}\cos{\phi}$. We have additionally chosen the angles $\phi$ and $\theta$ to display the lattice structure in the other two superradiant phases. For clarity, the selected values correspond to the case where the cavity amplitudes contribute equally to the total potential. The other points corresponding to equal distributions are associated with the same patterns but with a displacement of the lattice or with exchange between minima and maxima (see Fig.~\ref{plotSO}(c) and (d)). In general, other choices will have the same minima, but the landscape surrounding these points will not be as symmetric as in Fig.~\ref{plotSO}. In (a), we show the potential for the S1 phase, where cavity 1 is in the superradiant state. The atoms form a checkerboard pattern in the $x$-$y$ plane (not shown) but are free in the $z$ direction. Panel (b) corresponds to the S2 phase, where cavity 1 and 2 are superradiant. In this case, the atoms arrange in a rhomboidal lattice with lattice constant $2\lambda_{c}/\sqrt{3}$, being $\lambda_{c}$ the cavity wavelength. Finally, in (c) and (d), we present the potentials for the S3 phase, where all cavities are superradiant, for two situations where minima and maxima are exchanged. In both cases, the resulting structure is a hexagonal lattice, with lattice constant $2\lambda_{c}/3$ in (c) and $2\lambda_{c}/\sqrt{3}$ in (d). This set of qualitatively different self-organized structures is a consequence of the high complexity of the system, namely, the interference between the three different cavity modes. 

\section{Conclusion}
\label{Conc}

In this paper, we have studied the interaction between a gas of ultracold atoms and three single-mode optical cavities, in the presence of transverse pumping.  We showed that the $\mathbb{Z}_{2}$ symmetries associated with each cavity-atom coupling can be combined, not only into a $U(1)$ symmetry, but also into a rotational $SO(3)$ symmetry, which in the case of $n$ different single-mode resonators generalizes to an $SO(n)$ symmetry. Using the mean field approximation, we calculated the ground state phase diagram and obtained that spontaneous breaking of this symmetry led to a phase transition into a state where all cavities become superradiant, with the continuous manifold of degenerate ground states corresponding to different light field intensity distributions among the three cavities, which conserve the overall intensity present in the system. We found signatures of the $SO(3)$ symmetry breaking in the excitation spectrum, with the appearance of two Goldstone modes at the critical point (see Fig.~\ref{PlotExc}(c)), associated with the two possible excitations along the angular directions of the ground state manifold. We also studied the self-organization of the atoms, which resulted in a hexagonal lattice, whose precise periodicity depends on the specific realization of the symmetry breaking that occurs at the phase transition (see Figs.~\ref{plotSO}(c) and (d)).

In conclusion, our results demonstrate that rich phenomena emerge from multi-mode light-matter interacting systems. This is interesting from the perspective of quantum simulation, where efforts are being made towards the study of many-body systems with increasingly higher complexity \cite{Lev,Lev2,KeelingLevPRX}. Further interesting avenues include accounting for inter-cavity interactions, where coupling between order parameters allow to control the position of the phase boundaries \cite{EsslingerOrdPar}, and considering the out-of-equilibrium nature of the system, e.g.~the effects  quantum noise due to measurement back-action on the system dynamics \cite{DomokosPRL2010,DomokosPRA2011,DomokosBHM}, or the effects of photon losses on the steady state phase diagram, which have shown to wash out the presence of the continuous symmetry breaking \cite{Soriente}.

\section{Acknowledgements}
\label{Ack}

We are grateful to Nishant Dogra, Austen Lamacraft and Ulrich Schneider for fruitful and stimulating discussions.
E.I.R.C.~acknowledges support from the Winton Programme for the Physics of Sustainability and the UK Engineering and Physical Sciences Research Council (EPSRC) under Grant No.~EP/N509620/1. A.N.~acknowledges a University Research Fellowship from the Royal Society and additional support from the Winton Programme for the Physics of Sustainability.

\appendix

\section{Derivation of effective Hamiltonian \eqref{eq:er0002}}
\label{AppA}
The setup consists of a gas of ultracold two-level atoms forming a BEC, located at the intersection point of three high-finesse Fabry-P\'{e}rot cavities, lying in the $x$-$z$ plane, symmetrically aligned from each other and transversely pumped in the $y$ direction by a circularly polarized laser (see Fig.~\ref{FigModel}). We start by considering the single-particle Hamiltonian of a two-level atom interacting with the cavity modes and the external pump. In the rotating frame of the pump, the interaction Hamiltonian reads
\begin{equation}
\label{eq:erA01}
\hat{H}_{\textrm{int}}= -\hat{\vec{d}}^{\, \dagger} \cdot \hat{\vec{E}} - \hat{\vec{E}}^{\dagger} \cdot \hat{\vec{d}},
\end{equation}
where we have made use of the rotating-wave and dipole approximations. The dipole operator is defined $\hat{\vec{d}}=\vec{d} \hat{\sigma}_{-} e^{-i\omega_{p}t}$, with matrix element $\vec{d}=\langle g |\hat{\vec{x}}|e \rangle$, where $\hat{\vec{x}}$ is the position of the atom, $\hat{\sigma}_{-}=|g\rangle \langle e|$ is the lowering operator, and $|g \rangle$ and $|e \rangle$ are the ground and excited states of the atom, respectively. The electric field is defined by the linear combination of the three cavity fields and the external pump, yielding
\begin{equation}
\label{eq:erA02}
\hat{\vec{E}}= \sum_{l=1}^{3} E_{l}\vec{\epsilon}_{l} g_{l}(\vec{x})  \hat{a}_{l} e^{-i\omega_{p}t} + \frac{E_{p}}{2} \vec{\epsilon}_{p} g_{p}(\vec{x}) e^{-i\omega_{p}t},
\end{equation}
where $\omega_{p}$ is the pump frequency and $E_{l,p}$ and $\vec{\epsilon}_{l,p}$ are the field amplitudes and polarization vectors for cavity $l$ and the pump, respectively. The mode functions for the cavities and the pump are $g_{l}(\vec{x})\propto \cos{(\vec{k}_{c}^{(l)}\vec{x})}$ and $g_{p}(\vec{x}) \propto \cos{(\vec{k}_{p}\vec{x})}$, with the wave-vectors defined as in the main text. In the dispersive regime, where the driving is far detuned from the resonance frequency of the atom, the excited state can be eliminated using perturbation theory. This results in the dipole operator being proportional to the electric field $\hat{\vec{d}}=-\alpha_{s}\hat{\vec{E}}$ \cite{polarization}, where $\alpha_{s}\propto \Delta_{\textrm{at}}^{-1}$ corresponds to the scalar polarizability of the atoms, with $\Delta_{\textrm{at}}=\omega_{\textrm{at}}-\omega_{p}$ the atom-pump detuning, being $\omega_{\textrm{at}}$ the energy splitting of the two-level atom. This leads to

\begin{align}
\label{eq:erA03}
\hat{H}_{\textrm{int}}=& \sum_{l,l'=1}^{3}\alpha_{s} E_{l}E_{l'} g_{l}(\vec{x})g_{l'}(\vec{x})( \vec{\epsilon}_{l} \cdot\vec{\epsilon}_{l'}^{\, *}) \hat{a}^{\dagger}_{l'}\hat{a}_{l} \nonumber \\
& + \sum_{l=1}^{3} \frac{\alpha_{s} E_{l}E_{p}}{2}g_{l}(\vec{x})g_{p}(\vec{x}) \left[( \vec{\epsilon}_{l} \cdot\vec{\epsilon}_{p}^{\, *}) \hat{a}_{l} + \textrm{h.c.} \right] \nonumber \\
&+ \frac{\alpha_{s} E_{p}^{2}}{4}g_{p}(\vec{x})^{2}|\vec{\epsilon}_{p}|^{2}.
\end{align}
In general, the atomic polarization also has vectorial and tensorial components. Taking the atoms to be $^{87}$Rb with $F=1$ as the maximum eigenvalue  of the total angular momentum in the ground state manifold, the contribution from the vectorial component vanishes if we consider the case $m_{F}=0$ \cite{polarization}, being $m_{F}=-F,\dots,F$ the spin quantum number along the quantized axis. Furthermore, the tensorial component can be neglected in the typical frequency range used in these experiments \cite{EsslingerSpin}. We can simplify the Hamiltonian \eqref{eq:erA03} to obtain
\begin{align}
\label{eq:erA04}
\hat{H}_{\textrm{int}}=& \ U_{p} g_{p}(\vec{x})^{2} + \sum_{l=1}^{3} U_{l}g_{l}(\vec{x})^{2} \hat{a}^{\dagger}_{l}\hat{a}_{l}  \nonumber \\
& + \sum_{l=1}^{3} \eta_{l} g_{l}(\vec{x}) g_{p}(\vec{x}) \left( \xi_{l} \hat{a}_{l} + \xi^{*}_{l} \hat{a}_{l}^{\dagger}\right) , 
\end{align}
where we introduced the potential depths $U_{p}=\frac{\alpha_{s}E_{p}^{2}}{4}$ and $U_{l}=\alpha_{s}E_{l}^{2}$, the two-photon Rabi frequencies $\eta_{l}=\frac{\alpha_{s} E_{l} E_{p}}{2}$, the parameters $\xi_{l}=\vec{\epsilon}_{l} \cdot \vec{\epsilon}_{p}^{\, *}$, and used the definition $|\vec{\epsilon}_{l,p}|^{2}=1$.  We have neglected inter-cavity interactions, which is justified in the limit $E_{p} \gg E_{l}$. This results in a many-body Hamiltonian of the form ($\hbar=1$)
\begin{align}
\label{eq:erA05}
\hat{H} =& \sum_{l=1}^{3} (-\Delta_{l}) \hat{a}_{l}^{\dagger} \hat{a}_{l} + \int d\vec{x} \ \hat{\Psi}^{\dagger}(\vec{x}) \Bigg\{ \frac{\hat{p}_{x}^{2}+\hat{p}_{y}^{2}}{2m} \nonumber \\
&+ \sum_{l=1}^{3} \Bigg[  \eta_{l} \left( \xi_{l} \hat{a}_{l} + \xi^{*}_{l} \hat{a}_{l}^{\dagger}\right) \cos{(\vec{k}_{p}\vec{x})} \cos{(\vec{k}_{c}^{(l)}\vec{x})} \nonumber \\
&+  U_{l} \cos^{2}{(\vec{k}_{c}^{(i)} \vec{x})} \hat{a}_{l}^{\dagger} \hat{a}_{l}  \Bigg] + U_{p} \cos^{2}{(\vec{k}_{p} \vec{x})} \Bigg\}\hat{\Psi}(\vec{x}),
\end{align}
where $\Delta_{l}$ is the cavity-pump detuning as defined in the main text, $\hat{\Psi}(\vec{x})$ is the bosonic annihilation operator for the atomic field, $\hat{p}_{x,y}$ are the momentum operators, and $m$ is the mass of the atoms. The first term inside the curly brackets corresponds to the kinetic energy of the atoms. The second term is associated with light-matter interactions, where absorption or emission of a photon mediates transitions of the atoms between the BEC momentum state $|\vec{k}_{0}\rangle$ and twelve different momentum states $|\vec{k}^{(i)}=\pm(\vec{k}_{p} \pm \vec{k}^{(i)}_{c}) \rangle$. We focus on the case where $|\vec{k}_{c}^{(i)}|=|\vec{k}_{p}|=k$, which leads to the energies of the excited momentum states becoming degenerate $\omega=2\omega_{\textrm{rec}}$, with $\omega_{\textrm{rec}}=k^2/2m$ the recoil energy of the atoms. The third term is a dispersive shift of the cavity frequency due to the presence of the atoms and the last term corresponds to the periodic potential for the atoms generated by the pump. For simplicity, we have neglected the effects of short ranged interactions. We now consider the low-energy physics of the system and use the ansatz 
\begin{equation}
\label{eq:erA06}
\hat{\Psi}(\vec{x})=\frac{1}{\sqrt{V}}\hat{b}_{0} + \sum_{l=1}^{3} \frac{2}{\sqrt{V}}\cos{(\vec{k}_{p}\vec{x})} \cos{(\vec{k}_{c}^{(l)}\vec{x})} \hat{b}_{l},
\end{equation}
where $V$ is the volume of the system, and the operators $\hat{b}_{0,l}$ correspond to the bosonic annihilation operators of the momentum states defined in the main text. Inserting this ansatz into Eq.~\eqref{eq:erA05}, we obtain the effective Hamiltonian
\begin{equation}
\label{eq:erA07}
\hat{H}= \sum_{l=1}^{3} (-\Delta_{l}) \hat{a}_{l}^{\dagger} \hat{a}_{l} +  \omega \hat{b}_{l}^{\dagger} \hat{b}_{l} + \frac{\lambda_{l}}{\sqrt{N}} \left(\xi_{l}\hat{a}_{l} + \xi_{l}^{*} \hat{a}_{l}^{\dagger} \right) \left(\hat{b}_{l}^{\dagger} \hat{b}_{0} + \hat{b}_{0}^{\dagger} \hat{b}_{l} \right),
\end{equation}
where we absorbed the dispersive shift of the cavity in the definition of the detuning and we introduced the light-matter couplings $\lambda_{l}=\eta_{l}\sqrt{N}/2$. We define the cavity fields to be linearly polarized in the $x$-$z$ plane, with $\vec{\epsilon}_{1}=\hat{e}_{z}$, $\vec{\epsilon}_{2}=-\frac{\sqrt{3}}{2}\hat{e}_{x}+\frac{1}{2}\hat{e}_{z}$ and $\vec{\epsilon}_{3}=\frac{\sqrt{3}}{2}\hat{e}_{x}+\frac{1}{2}\hat{e}_{z}$. To ensure the possibility of realizing symmetrical coupling between the cavities we choose the pump to be circularly polarized $\vec{\epsilon}_{p}=e^{-i\frac{\pi}{2}}\hat{e}_{x}+\hat{e}_{z}$, leading to $(\xi_{1},\xi_{2},\xi_{3})=(1,e^{i\frac{\pi}{3}},e^{-i\frac{\pi}{3}})$. These factors can then be removed by performing a set of unitary transformations of the form $\hat{a}_{l} \rightarrow \hat{a}_{l}/\xi_{l}$, which yield the effective Hamiltonian presented in Eq.~\eqref{eq:er0002}. 

\section{Calculation of the energy spectrum}
\label{AppB}
We obtain the energy spectrum of the Hamiltonian \eqref{eq:er0002} following the approach used in Refs.~\cite{Emary,BaksicCiutiPRL}. First, we displace the operators by their expectation values $\hat{a}_{i} \rightarrow \hat{c}_{i} + \alpha_{i}$, $\hat{b}_{i} \rightarrow \hat{d}_{i} - \beta_{i}$. Considering the thermodynamic limit $N \rightarrow \infty$, we expand the Hamiltonian up to order $N^{0}$ using that ($\alpha_{i}$,$\beta_{i}$) $\propto \sqrt{N}$, where the leading order term corresponds to the mean-field energy $E_{\textrm{MF}}$ defined in Eq.~\eqref{eq:er0008}. Imposing the terms linear in $(\hat{c}_{i},\hat{d}_{i})$ to be vanishing leads to the mean-field solutions obtained in Sec.~\ref{MFPD}. We are thus left with a bilinear Hamiltonian
\begin{align}
\label{eq:erB01}
\hat{H}_{\textrm{bil}} =& \sum_{i=1}^{3} \Bigg\{ (-\Delta) \hat{c}^{\dagger}_{i} \hat{c}_{i} \left[ \omega + \left( \sum_{j} \frac{2\lambda_{j}}{\sqrt{kN}} \alpha_{j}\beta_{j} \right) \right] \hat{d}_{i}^{\dagger} \hat{d}_{i} \nonumber \\
&  + \frac{\lambda_{i}}{\sqrt{kN}} \sum_{j=1}^{3} \alpha_{i} \beta_{j} \left(\hat{d}^{\dagger}_{i} \hat{d}^{\dagger}_{j}  + \hat{d}^{\dagger}_{i} \hat{d}_{j} + \hat{d}^{\dagger}_{j} \hat{d}_{i} +\hat{d}_{j} \hat{d}_{i}\right) \nonumber \\
&+ \frac{\lambda_{i}}{2 k\sqrt{kN}} \sum_{j,l=1}^{3} \alpha_{i} \beta_{i} \beta_{j} \beta_{l} \left( \hat{d}_{j}^{\dagger} + \hat{d}_{j} \right) \left( \hat{d}_{l}^{\dagger} + \hat{d}_{l} \right) \nonumber \\
&+\lambda_{i} \left( \hat{c}_{i}^{\dagger} + \hat{c}_{i}\right) \sum_{j=1}^{3} \left( \delta_{ij} \sqrt{\frac{k}{N}} -  \frac{\beta_{i} \beta_{j}}{\sqrt{kN}} \right) \left( \hat{d}_{j}^{\dagger} + \hat{d}_{j}\right)  \Bigg\} ,
\end{align}
where $k=N-\sum_{i} \beta_{i}$. Inserting the different solutions for $\alpha_{i}$ and $\beta_{i}$ described in Sec.~\ref{MFPD} yields the Hamiltonian in each different phase. Given the bilinear nature of \eqref{eq:erB01}, the normal modes and energy spectrum of the system in each phase are straightforwardly obtained by performing a Bogoliubov transformation.

\bibliography{library}

\end{document}